\documentclass[debug]{rmaa}

\usepackage{paralist}
\usepackage{longtable}
\usepackage{lipsum} % just for dummy text- not needed for a longtable
\usepackage{subfigure}
\usepackage{psfrag,color}

\usepackage[latin1]{inputenc}

\title{Study of an intermediate age open cluster IC 1434 using ground-based imaging and Gaia DR2 astrometry} 

\author{
	Y.H.M. Hendy,\altaffilmark{1}
  and D. Bisht$^{*}$\altaffilmark{2}}

\altaffiltext{1}{Astronomy Department, National Research Institute of Astronomy and Geophysics (NRIAG), Helwan, Cairo, Egypt.}

\altaffiltext{2}{Key Laboratory for Researches in Galaxies and Cosmology, University of Science and Technology of China, Chinese
          Academy of Sciences, Hefei, Anhui, 230026, China.}

\shortauthor{Hendy, Bisht}
\shorttitle{Study of IC 1434 in Gaia era}

\fulladdresses{
\item Y.H.M. Hendy: Astronomy Department, National Research Institute of Astronomy and Geophysics (NRIAG), Helwan, Cairo, Egypt (y.h.m.hendy@gmail.com).

\item D. Bisht: Key Laboratory for Researches in Galaxies and Cosmology, University of Science and Technology of China, Chinese Academy of Sciences,
      Hefei, Anhui, 230026, China (dbisht@ustc.edu.cn).
}

\listofauthors{W. J. Henney, A. Collaborator, \& L. Author}
\indexauthor{Hendy, Y.H.M.}
\indexauthor{Bisht, D.}

\abstract{We present a detailed photometric and kinematical analysis of  poorly studied open cluster IC 1434 using CCD VRI,  APASS, and
Gaia~DR2 database for the first time. By determining the membership probability of stars, we identified the 238 most probable
members with a probability higher than $60\%$ by using proper motion and parallax data as taken from the Gaia~DR2 catalog. The mean proper motion
of the cluster is obtained as $\mu_{x}=-3.89\pm0.19$ and $\mu_{y}=-3.34\pm0.19$ mas yr$^{-1}$ in both the directions of right ascension
and declination. The radial distribution of member stars provides cluster extent as 7.6 arcmin. We have estimated the interstellar
reddening (E(B-V)) as 0.34 mag using the transformation equations from literature. We obtained the values of cluster age and distance are $631\pm73$ Myr
and $3.2\pm0.1$ Kpc.}
\resumen{.}

\addkeyword{Open star cluster IC 1434}
\addkeyword{Color Magnitude Diagram}
\addkeyword{Astrometry}
\begin{document}
% Typeset article header
\maketitle

\section{INTRODUCTION}
\label{sec:intro}

%%%%%%%%%%%%%%%%
\begin{figure*}
\begin{center}
\includegraphics[width=10.5cm, height=10.5cm]{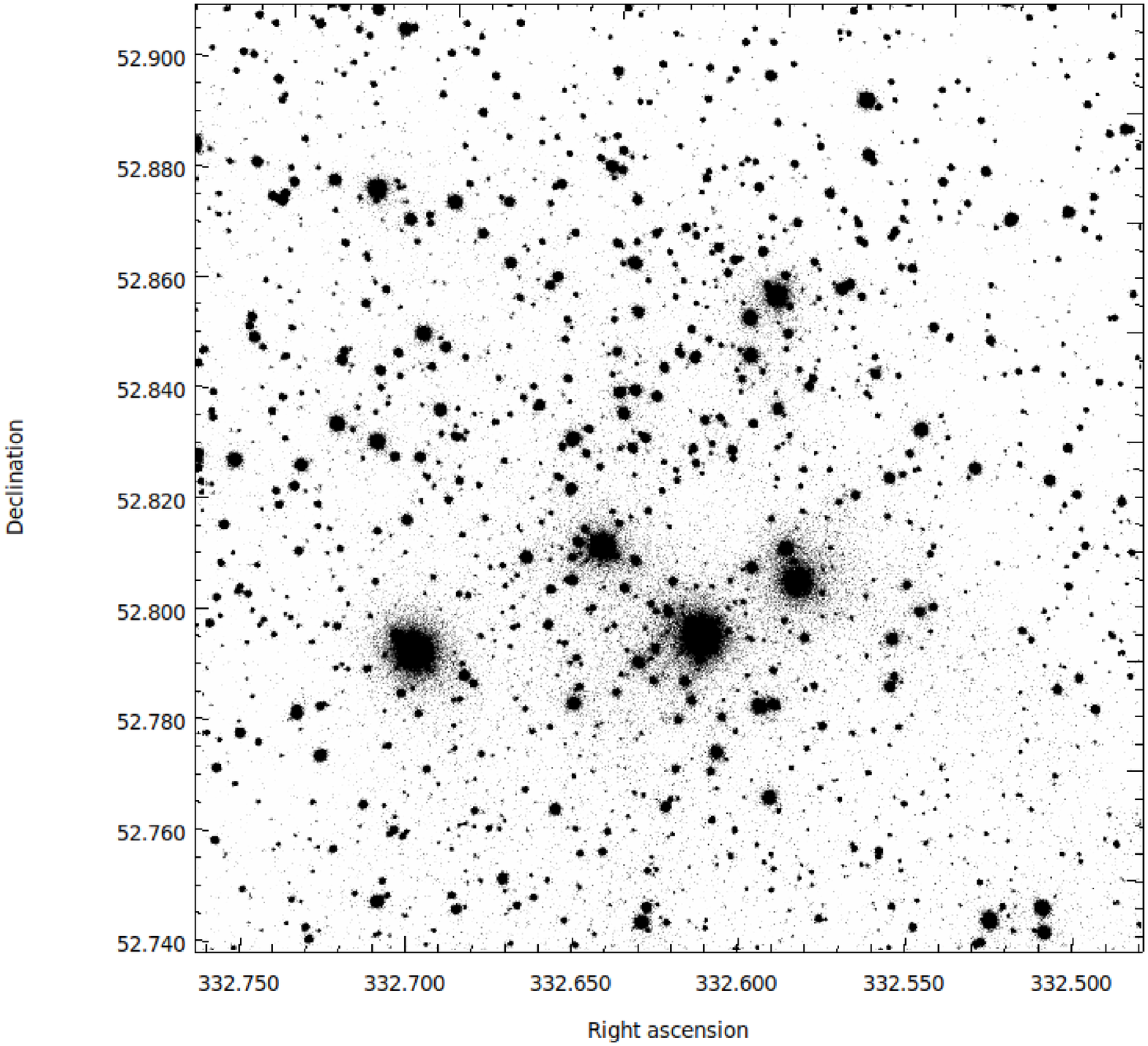}
\caption{Finding chart of the stars in the field of IC 1434. Filled circles of different sizes represent
the brightness of the stars. The smallest size denotes stars of $V\sim20$ mag.}
\label{id} 
\end{center}
\end{figure*}
%%%%%%%%%%%%%%%%
Open clusters (OCs) are important tools to probe the Galactic disk properties (see, e.g., Friel 1995). OCs are very advantageous
to interpret many queries regarding the assessment of chemical abundance gradients in the disk (see, e.g., Twarog,
Ashman \& Anthony-Twarog 1997; Chen, Hou \& Wang 2003), Galactic structure and evolution (e.g., Janes \& Adler 1982;
Janes \& Phelps 1994), interactions between thin and thick disks (e.g., Sandage 1988), as well as the theories of stellar
formation and evolution (e.g., Meynet, Mermilliod \& Maeder 1993; Phelps \& Janes 1993). It's not an easy task to segregate
cluster members from field stars considering OCs are generally projected against the Galactic disc stars. The second data release
(DR2) (Gaia Collaboration et al. 2016a) contains 1.7 billion sources and was made public on 2018 April 24 (Jordi et al. 2010;
Gaia Collaboration et al. 2016a,b; Salgado et al. 2017). The Gaia~DR2 data contains photometric magnitudes in three bands
($G$, $G_{BP}$, and $G_{RP}$) and astrometric data at the sub-milliarcsecond level along with the parallax (Gaia Collaboration
et al. 2018a). Gaia data has been used recently by many authors to estimate the membership probability of stars lying towards
the cluster regions
(Cantat-Gaudin et al. 2018; Gao 2018; Rangwal et al. 2019; Bisht et al. (2019, 2020a, 2020b)).

The open cluster IC 1434 ($\alpha_{2000} = 22^{h}10^{m}30^{s}$, $\delta_{2000}=52^{\circ} 50^{\prime} 00^{\prime\prime}$;
$l$=99$^\circ$.937, $b$=-2$^\circ$.700) is located in the second Galactic quadrant. Tadross (2009) analyzed this object
using 2MASS and NOMAD data sets. He obtained the age, interstellar reddening ($E(B-V)$),  and distance of this object as 0.32 Gyr,
0.66 mag, and $3035\pm140$ pc, respectively. In this paper, our main goal is to accomplish a deep and precise analysis of an
intermediate-age open cluster IC 1434 using CCD $VRI$, APASS, and Gaia DR2 data.

%\textbf{Mass function study of OCs is very helpful to understand the star formation process (Kroupa 2001; Elmegreen 1999; Richtler 1994).
%Colour-magnitude diagram of clusters can be used to derive luminosity function (LF) and mass function (MF).
%Considerable work has been accomplished by many researchers on mass function studies of clusters (Phelps \& Janes 1993;
%Piskunov et al. 2004, Scalo 1986; Yadav \& Sagar 2002, 2004a). Our understanding for the universality of mass
%function is still like a puzzle (Elmegreen 2000). In this paper, one of our main aim is to discuss mass function of IC 1434
%using good quality data.}

The layout of the paper is as follows. A brief description of data used, data reduction, and calibration are
described in Section 2. Section 3 deals with the study of proper motion and determination of membership probability
of stars. The structural properties and derivation of fundamental parameters using the most probable cluster members
have been carried out in Section 4. The conclusions are presented
in Section 5.

\section{Observations and calibration of CCD data}

The $VRI$ CCD photometric observations of IC 1434 was carried out using the 74-inch Kottamia astronomical observatory
(KAO) of NRIAG in Egypt. Images were collected using a $2k\times2k$ CCD system. The observations were taken at the Newtonian
focus with a field area of $10^{\prime}\times10^{\prime}$ and a pixel scale of 0$^{\prime\prime}$.305 pixel$^{-1}$
on $8^{th}$ November 2013. The read-out noise was 3.9 e$^{-}$/pixel. Observations were organized in several short
exposures with the air mass ranges of 1.32-1.62 in each of the filter as described in Table~\ref{log}. All CCD frames
observed with two amplifiers, which treated for overscan, bias, and flat field corrections using an IRAF's code written
by one of the authors (Y.H.M. Hendy), see Tadross et al. (2018). To perform the photometry, we have used a DAOPHOT package
on IRAF (Stetson 1987, 1992). The data reduction procedure has been well explained by Bisht et al. (2019). The identification
chart for IC 1434 based on our $V$-band observations is shown in Fig.~\ref{id}.

To obtain the instrumental magnitudes of stars in the observed field, we used the point spread function of Stetson (1987). We have transformed the pixel coordinates (X and Y) into the right ascension and declination using the astrometry website
(https://nova.astrometry.net/).\\

%%%%%%%%%%%%%%%%
\begin{figure*}
\begin{center}
\includegraphics[width=8.5cm, height=10.5cm]{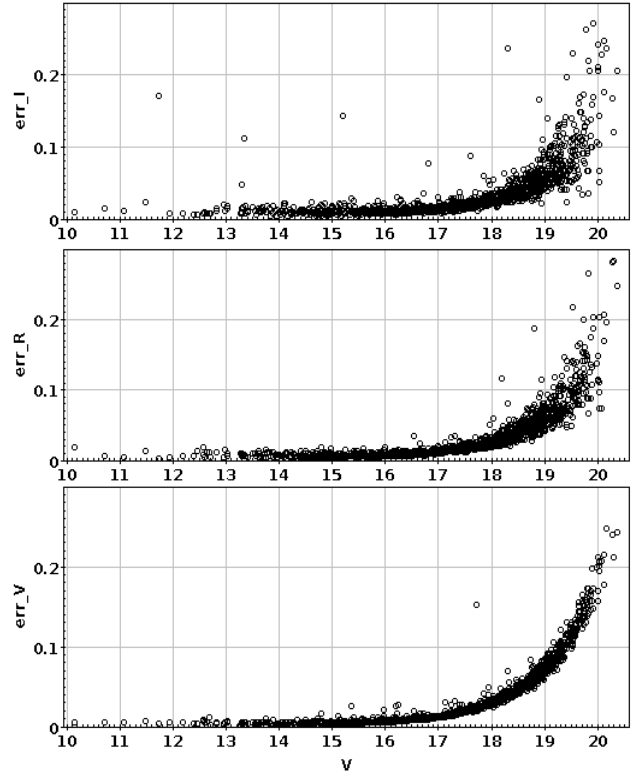}
\caption{Photometric errors in $V$, $R$, and $I$ bands against $V$ magnitude.}
\label{v_error} 
\end{center}
\end{figure*}
%%%%%%%%%%%%%%%%

\begin{table}
\begin{center}
\caption{Log of observations, with dates and exposure times for each passband.}
\vspace{0.5cm}
\begin{tabular}{ccc}
\hline\hline
Band  &Exposure Time &Date\\
&(in seconds)   & \\
\hline\hline
$V$&120$\times$3, 60$\times$1&8$^{th}$ November 2013 \\
$R$&120$\times$2, 60$\times$1&,,\\
$I$&120$\times$3, 60$\times$1&,,\\
\hline\hline
\end{tabular}
\label{log}
\end{center}
\end{table}
%%%%%%%%%%%%%%%%
\begin{figure*}
\begin{center}
\includegraphics[width=8cm, height=6cm]{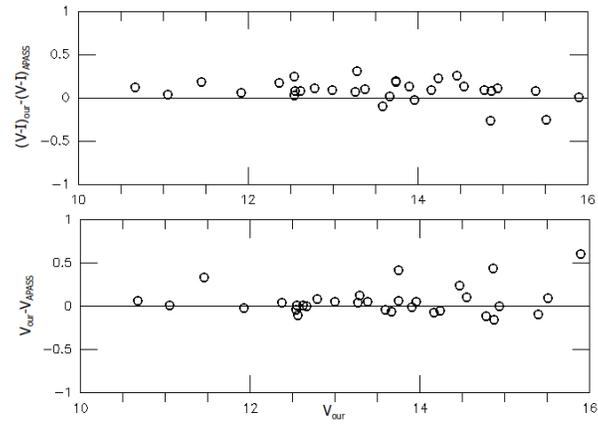}
\caption{Differences between measurements presented in the APASS catalog and this study
for $V$ magnitude and ($V$-$I$) colors. Zero difference is indicated by the solid line}
\label{comparison} 
\end{center}
\end{figure*}
%%%%%%%%%%%%%%%%
%%%%%%%%%%%%%%%%
\begin{figure*}
\begin{center}
\includegraphics[width=8.5cm, height=10.5cm]{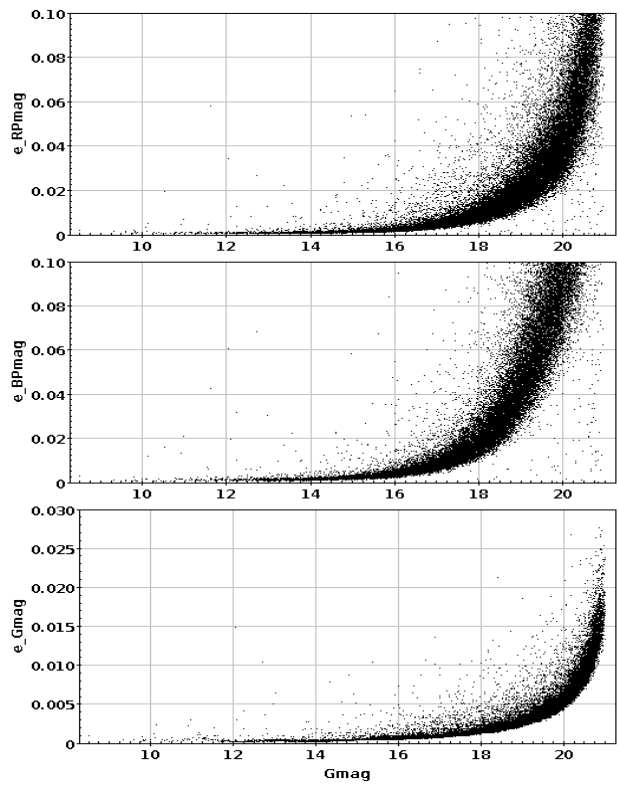}
\caption{Photometric errors in Gaia bands ($G$, $BP$, and $RP$) against $G$ magnitude.}
\label{gband_error} 
\end{center}
\end{figure*}
%%%%%%%%%%%%%%%%

%%%%%%%%%%%%%%%%
\begin{figure*}
\begin{center}
\includegraphics[width=8.5cm, height=14.5cm]{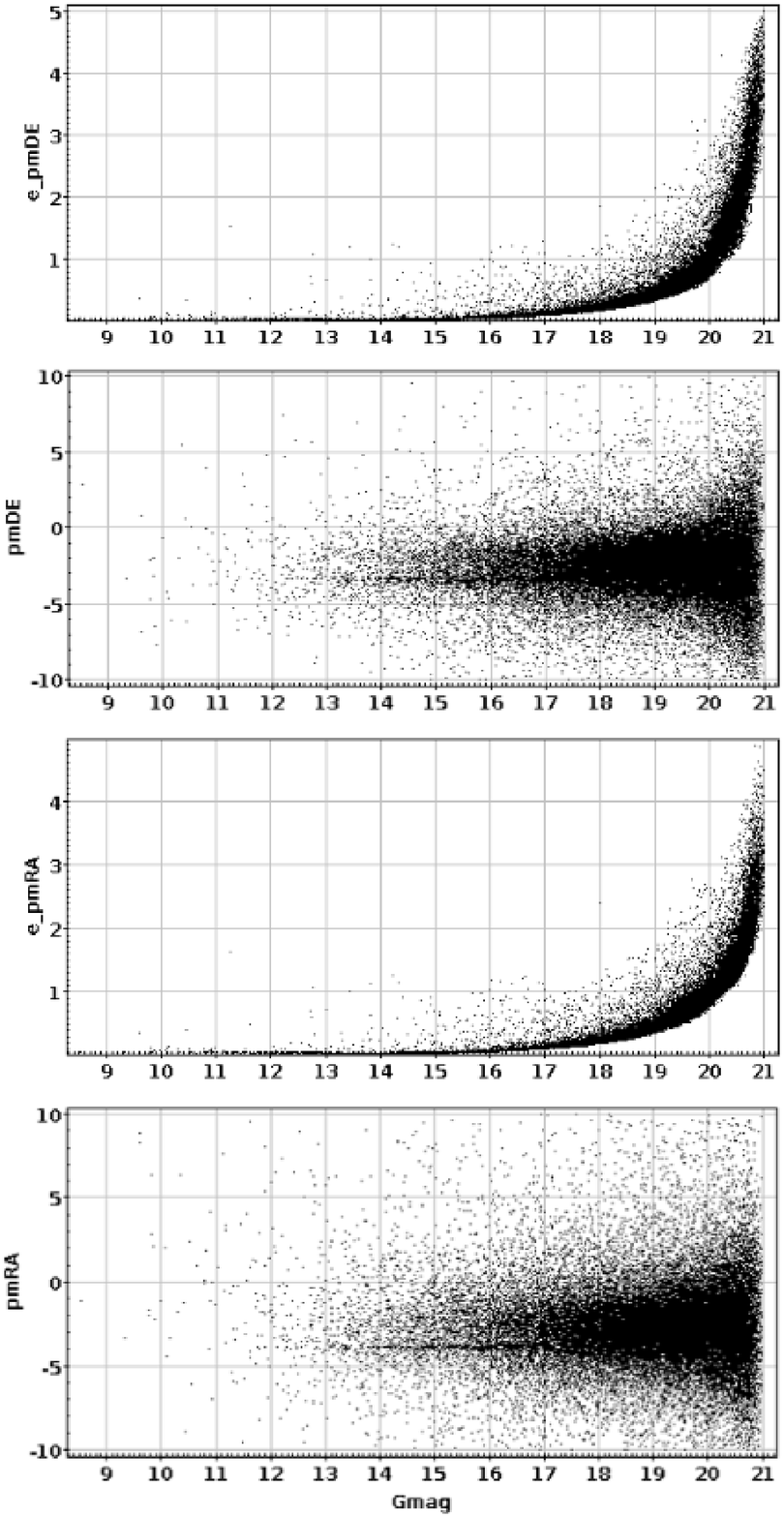}
\caption{The plot of proper motions and their errors versus $G$ magnitude.}
\label{pm_error} 
\end{center}
\end{figure*}
%%%%%%%%%%%%%%%%
To transform the $VRI$ instrumental magnitudes into Johnson and Kron-Cousin standard magnitudes, we have used the
photometric data available in $V$-band from Maciejewski \& Niedzielski (2008) and $RI$-bands from the USNO-B1.0 catalog (Monet et al. 2003).

The transformation equations for converting the instrumental magnitude in to standard magnitude, are as follows:

\begin{center}
   $V= V_{ins}-(0.14\pm0.039)(V-R)+21.18\pm0.031$\\

   $V-R= (1.06\pm0.035)(V-R)_{ins}+0.56\pm0.022$\\

   $V-I= (0.90\pm0.036)(V-I)_{ins}+0.85\pm0.027$\\
\end{center}

The respective errors in zero points and color coefficients are $\sim$0.03 mag shown in the above transformation equations.
The internal errors derived from DAOPHOT are plotted against $V$ magnitude shown in Fig.~\ref{v_error}. This figure shows that
the average photometric errors are $\le$ 0.02 mag at $V\sim18^{th}$ mag, while it is $\le$ 0.1 mag at $V\sim19^{th}$ mag.\\ 

To compare the photometry, we have cross-identified the stars in our observed data with the American Association of Variable Star
Observers (AAVSO) Photometric All-Sky Survey (APASS) DR9 catalog. We have assumed that stars are accurately matched if the
difference in position is less than 1 arcsec and in this way, we have found 32 common stars accordingly. The APASS survey is cataloged
in five filters $B$, $V$, $g$, $r$, and $i$. The range of magnitude in the $V$
band is from 7 to 17 mag (Heden \& Munari 2014).
The DR9 catalog covers  almost about 99 $\%$
of the sky (Heden et al. 2016). To obtain the Cousins $I$ band using Sloan $ri$ photometric bands from the APASS catalog
($I_{APASS}$ = $i$-(0.337 $\pm$ 0.191) ($r$-$i$)-(0.370 $\pm.041$)), we have adopted the method given by Tadross \& Hendy (2016).

The difference indicates that present $V$ and ($V$-$I$) measurements are in fair agreement with those stars given in
the APASS catalog. The comparable difference is found as 0.07 in the $V$ band and 0.08 in ($V$-$I$) as shown in Fig.~\ref{comparison}.

\subsection{Gaia DR2}

Gaia~DR2 (Gaia Collaboration et al. 2018b) database within a 20$^{\prime}$ radius of the cluster is used for the astrometric analysis.
This data consists of positions on the sky $(\alpha, \delta)$, parallaxes and proper motions
($\mu_{\alpha} cos\delta , \mu_{\delta}$) with a limiting magnitude of $G=21$ mag. The errors in photometric
magnitudes ($G$, $G_{BP}$, and $G_{RP}$) with G mag are shown in Fig \ref{gband_error}. In this figure, we find the mean errors in G the band is $\sim$ 0.01 mag while the mean errors in $G_{BP}$ and $G_{RP}$ bands are $\sim$ 0.1 mag at 20 mag. The proper
motions with their respective errors are plotted against $G$ magnitude is shown in Fig.~\ref{pm_error}. The uncertainties in the
corresponding proper motion components are $\sim$ 0.06 mas $yr^{-1}$ (for $G\le15$ mag), $\sim$0.2 mas $yr^{-1}$ (for $G\sim17$ mag), and $\sim$1.2 mas $yr^{-1}$ (for $G\sim20$ mag).\\

\section{Proper motion study and membership probabilities of stars}

%%%%%%%%%%%%%%%%
\begin{figure*}
\begin{center}
\includegraphics[width=7.5cm, height=7cm]{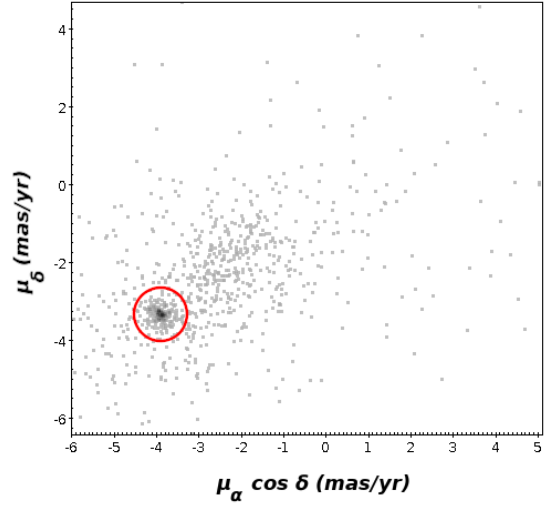}
\caption{Vector point diagram, circle defines the cluster region of IC 1434 within the radius of 0.6 mas/yr.}
\label{vpd} 
\end{center}
\end{figure*}
%%%%%%%%%%%%%%%%

%%%%%%%%%%%%%%%%
\begin{figure*}
\begin{center}
\includegraphics[width=8.5cm, height=8.5cm]{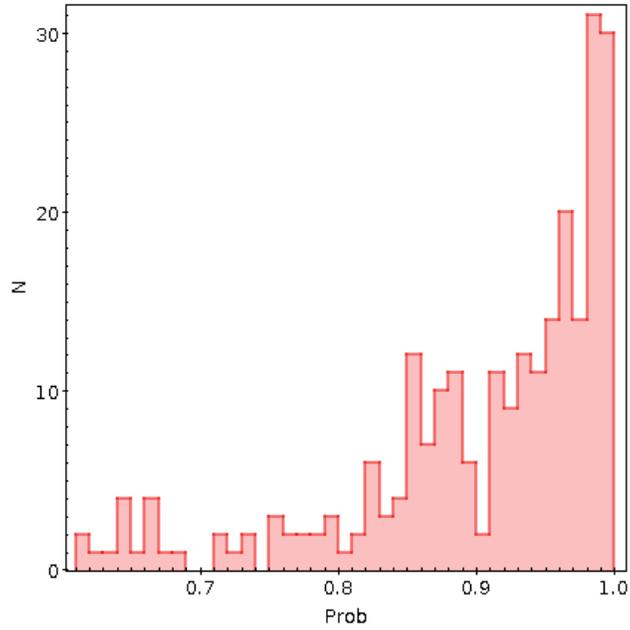}
\caption{A membership probability histogram of stars for IC 1434. We considered stars with a probability $\ge$ 0.6 as cluster members.}
\label{prob} 
\end{center}
\end{figure*}
%%%%%%%%%%%%%%%%

%%%%%%%%%%%%%%%%
\begin{figure*}
\begin{center}
\includegraphics[width=5cm, height=5cm]{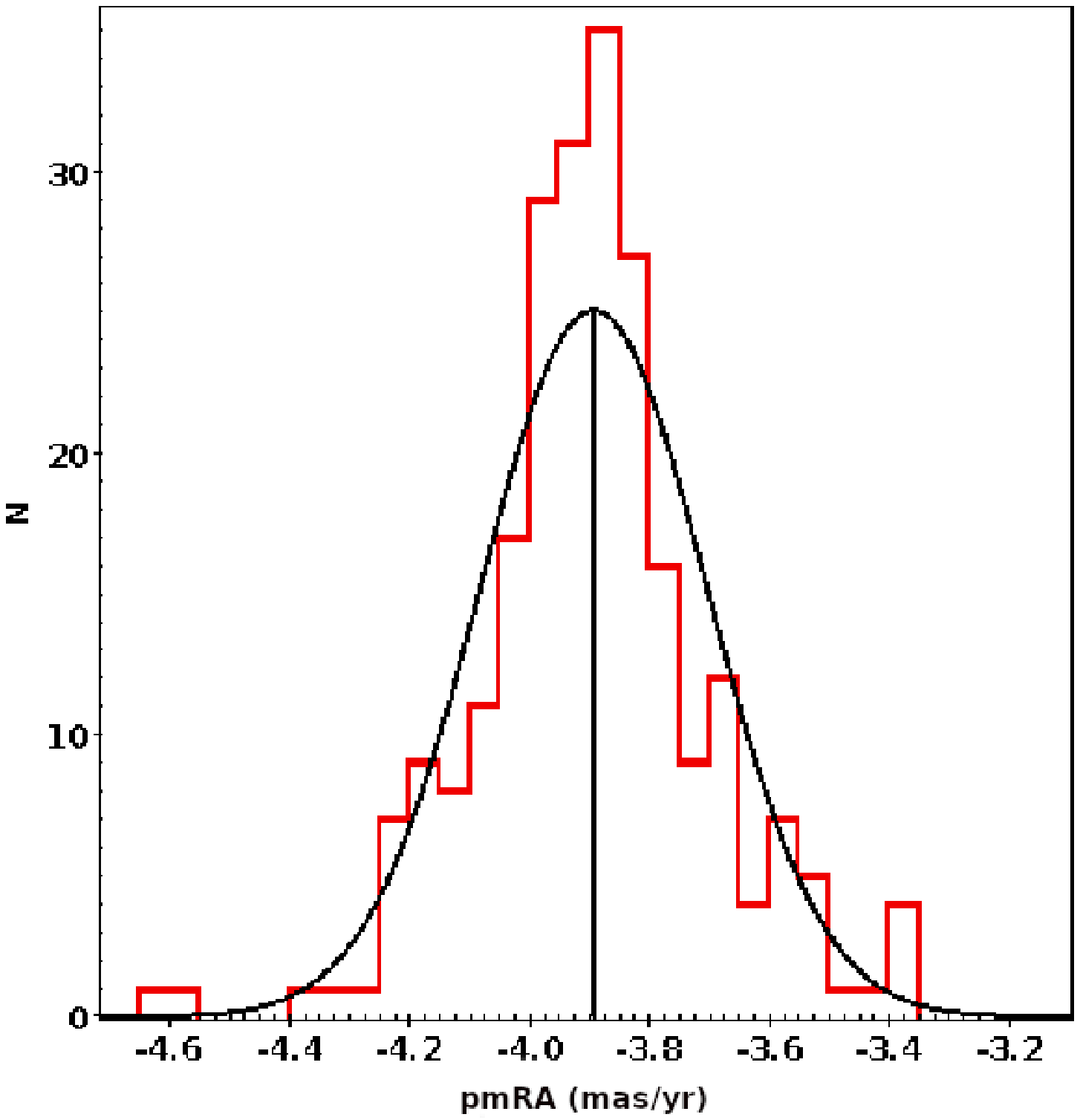}\\
%\vspace{-0.5cm} 
\includegraphics[width=5cm, height=5cm]{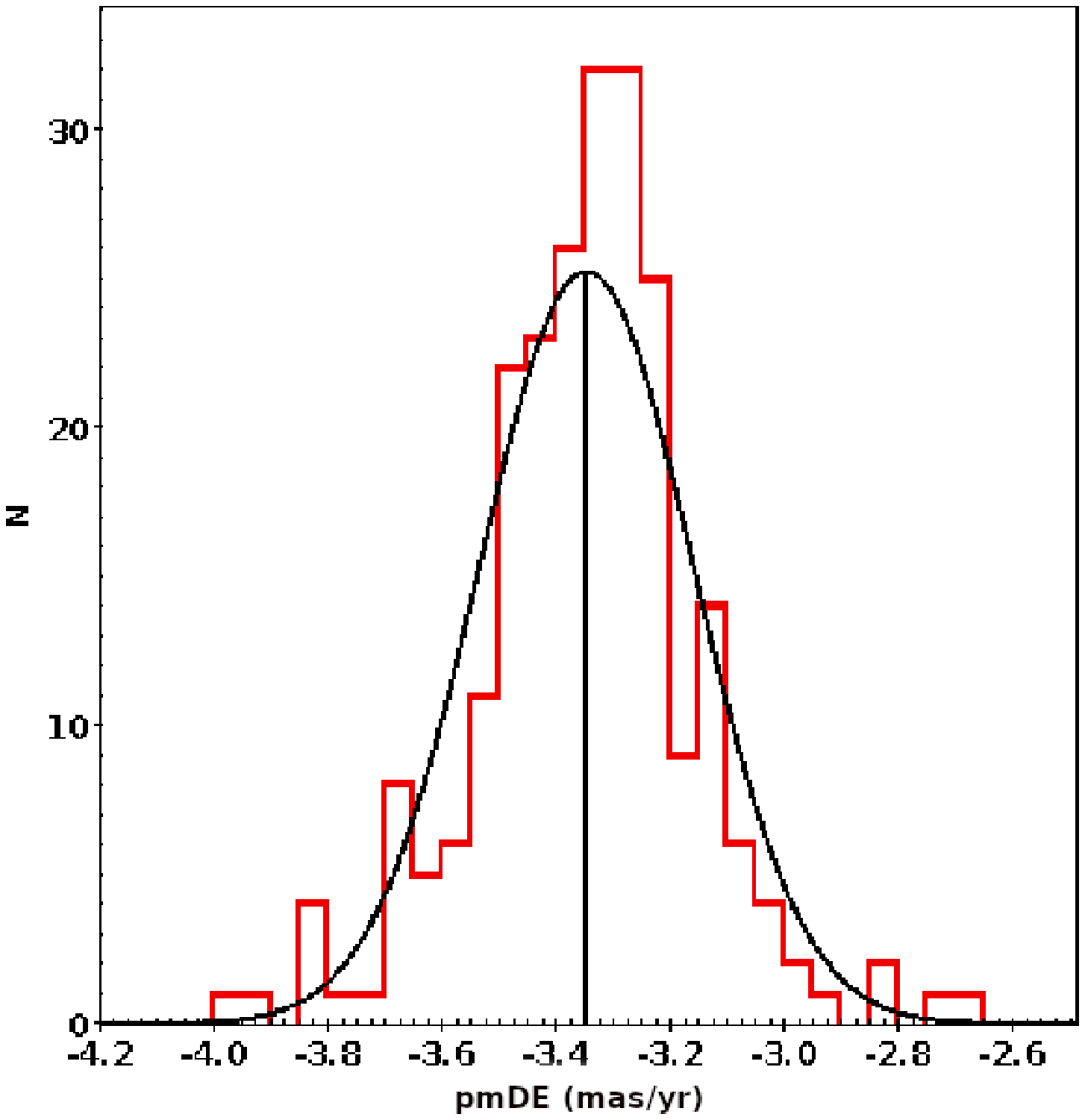}\\
%\vspace{-0.5cm} 
\includegraphics[width=5cm, height=5cm]{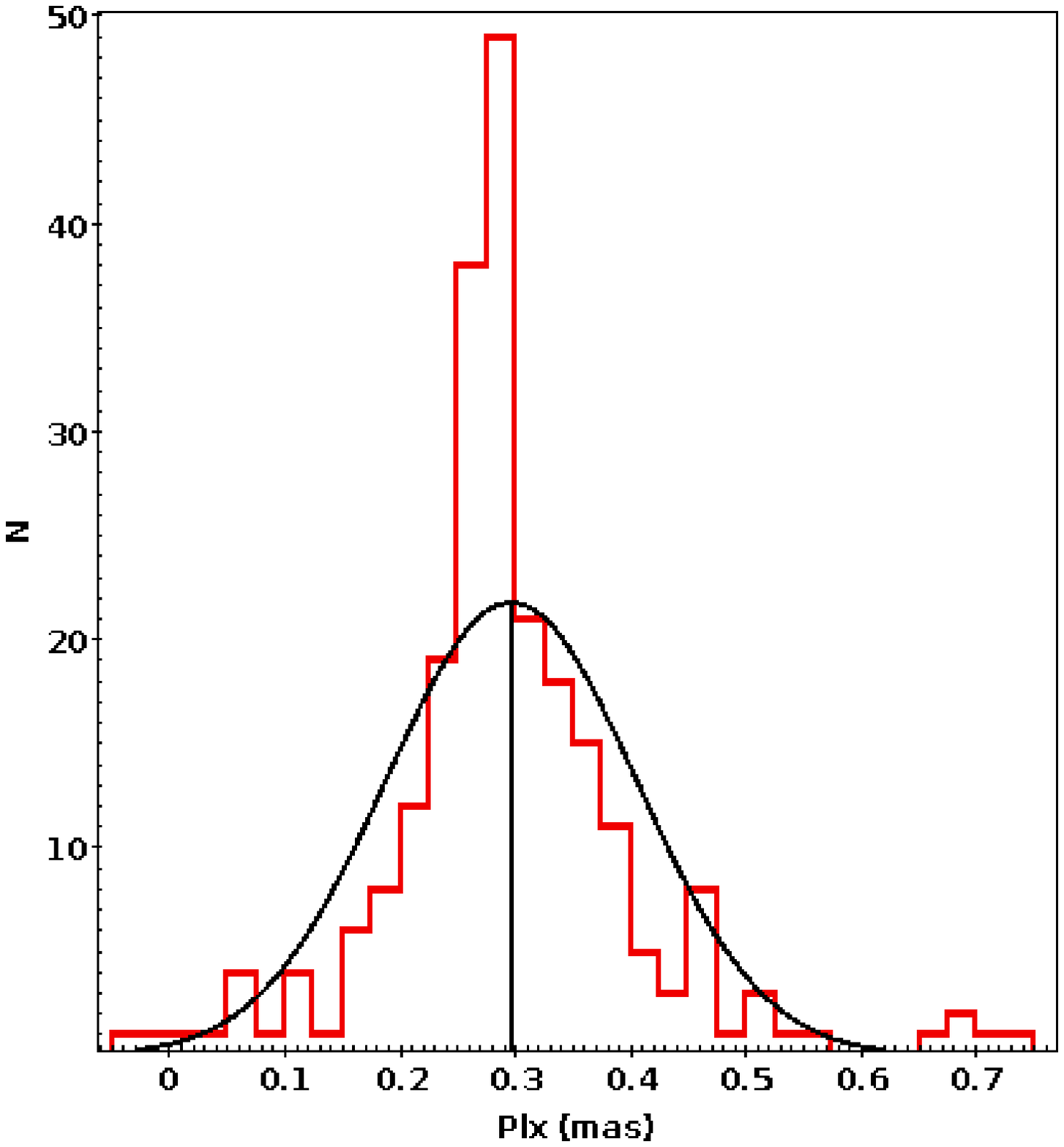}
\caption{Histogram to determine mean values of proper motions in RA and DEC directions (top and middle panels). Histogram to find mean parallax (bottom panel). The Gaussian function fits to the central bins provide the mean values in PMs and parallax shown in each panel.}
\label{pm} 
\end{center}
\end{figure*}
%%%%%%%%%%%%%%%%

%%%%%%%%%%%%%%%%
\begin{figure*}
\begin{center}
%\includegraphics[width=11cm, height=4cm]{G_pmra_2.ps}\\
%\vspace{-0.5cm} 
%\includegraphics[width=11cm, height=4cm]{G_pmde_2.ps}\\
%\vspace{-0.5cm} 
%\includegraphics[width=11cm, height=4cm]{G_plx_2.ps}
\includegraphics[width=11cm, height=12cm]{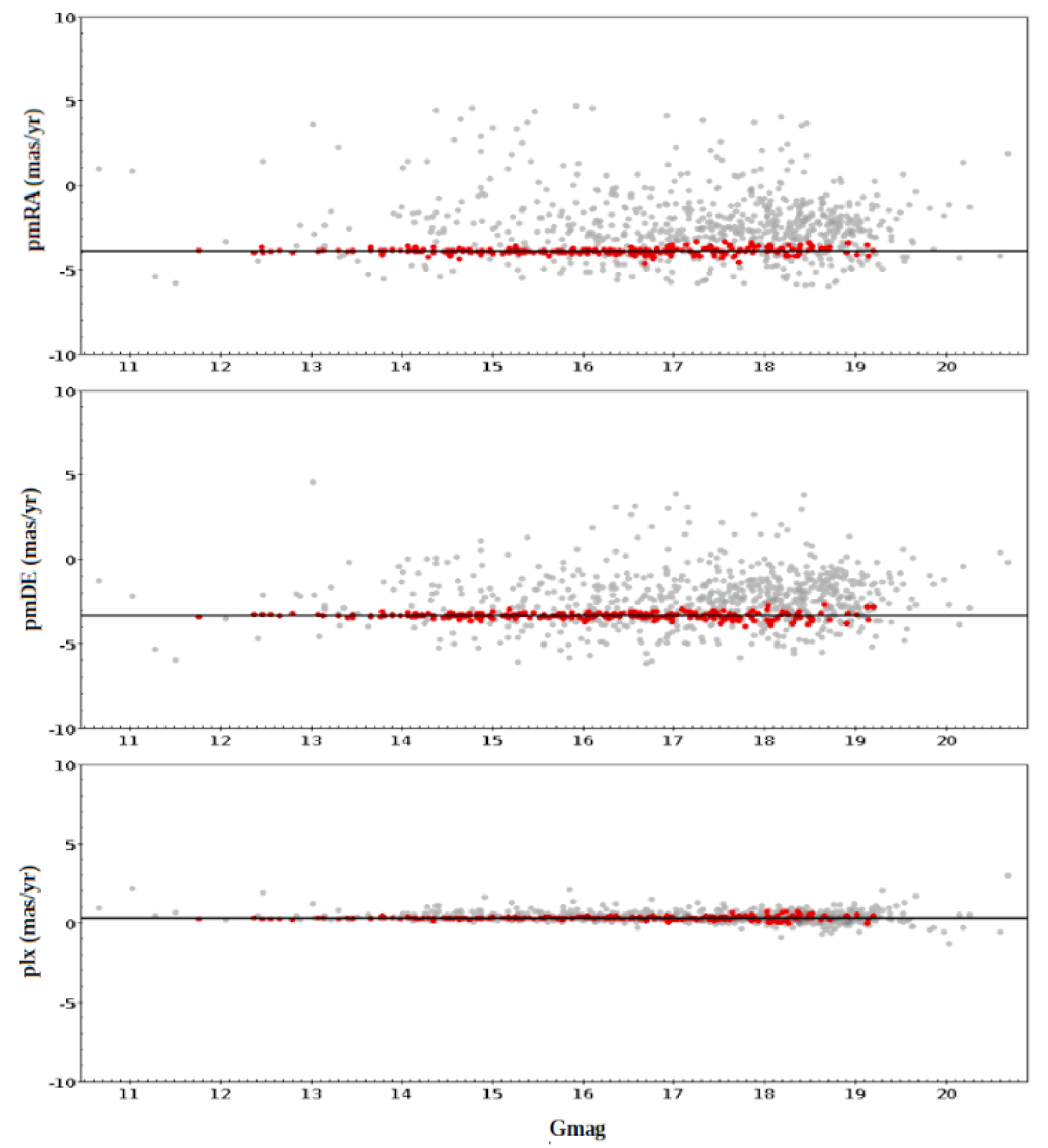}
\caption{Proper motion components and parallax distribution of most probable members (red dots) \& all stars (gray dots) against $G$ band magnitude. The horizontal line indicates the mean value of PMs and parallax.}
\label{g_mp} 
\end{center}
\end{figure*}
%%%%%%%%%%%%%%%%

The proper motion of a cluster is a change in its angular position with time as seen from the center of mass of the solar system.
Proper motions play an influential role to eliminate non-members from the cluster's main sequence (Yadav et al. 2013; Bisht et al. 2020a).
We have cross-matched our observational data in VRI bands and data from the Gaia DR2 catalog.
A circle of 0.6 $mas/yr$ around the cluster center in the VPD characterizes our membership criteria as shown in Fig.~\ref{vpd}.
The chosen radius in VPD is a compromise between losing members with poor proper motions and the inclusion of field region stars. 

The OCs are highly contaminated by a large number of foreground/background stars especially towards the fainter end of the main sequence.
Vasilevskis et al. (1958) have set up a
mathematical model to obtain membership probabilities of stars using proper motion data. A revised technique was developed
by Stetson (1980) and Zhao \& He (1990) to check the membership of stars in OCs based on proper motions. To find the membership
probability of stars towards the region of IC 1434, we have adopted the criteria as discussed by Kharchenko et al. (2004). This
method has been previously used by Bisht et al. (2018) for OCs Teutsch 10 and Teutsch 25. Hendy (2018) also has been used this method for
open cluster FSR 814. The kinematical probability of stars is expressed as:\\

%\begin{equation}
~~~~~~~~~~~~~~~~~~~~~~~~~~~~$p_{k}=e^{[-0.25({\frac{(\mu_{x}-\overline{\mu_{x}})^2}{\sigma_{x}^2}}+\frac{(\mu_{y}-\overline{\mu_{y}})^2}{\sigma_{y}^2})]}$\\
%\end{equation}

where $\sigma_{x}^2=\sigma_{\mu_x}^2+\sigma_{\overline{\mu_x}}^2$ and  $\sigma_{y}^2=\sigma_{\mu_y}^2+\sigma_{\overline{\mu_y}}^2$. Here $\mu_x$ and $\mu_y$ are the proper motion of a particular star, while $\sigma_{\mu_x}$ and $\sigma_{\mu_y}$ are the corresponding errors.
The ${\overline{\mu_x}}$ and ${\overline{\mu_y}}$ are the mean value of proper motions,  while
$\sigma_{\overline{\mu_x}}$ and $\sigma_{\overline{\mu_y}}$ are their corresponding standard deviations. 
Using the above method, we have identified possible members of IC1434 if the membership probability of stars higher
than $60\%$ which can be seen in Fig. 7.

We have used only probable cluster members to estimate the mean value of proper motions and parallax of IC 1434. We have fitted Gaussian profile to the construct histograms shown in Fig.~\ref{pm}. We obtained the mean-proper motion of IC 1434 $-3.89\pm0.19$ and $-3.34\pm0.19$ mas yr$^{-1}$ in RA and DEC directions, respectively. We determined the mean value of parallax $0.30\pm0.11$ mas. Our findings are very close with the values given by Cantat-Gaudin et al. (2018). 

Finally, we have considered a star as the most probable member if it lies within 0.6 mas/yr radius in VPD, having a membership probability higher
than $60\%$ and a parallax within $3\sigma$ from the mean parallax of the cluster IC 1434. In Fig.~\ref{g_mp}, we have plotted the
proper motions and parallax distribution of most probable members as denoted by red dots \& all observed stars as denoted by gray dots
against $G$ magnitude. In this figure, the horizontal solid line indicates the mean value of proper motions and parallax.

%In Fig.~\ref{vpd}, we have shown the vector point diagram and their corresponding $G (BP-RP)$ and $V (V-R)$ %color magnitude diagrams.

%obtained 238 most probable members 

\section{Structural properties of IC 1434}

%%%%%%%%%%%%%%%%
\begin{figure*}
\begin{center}
\includegraphics[width=13.5cm, height=7.5cm]{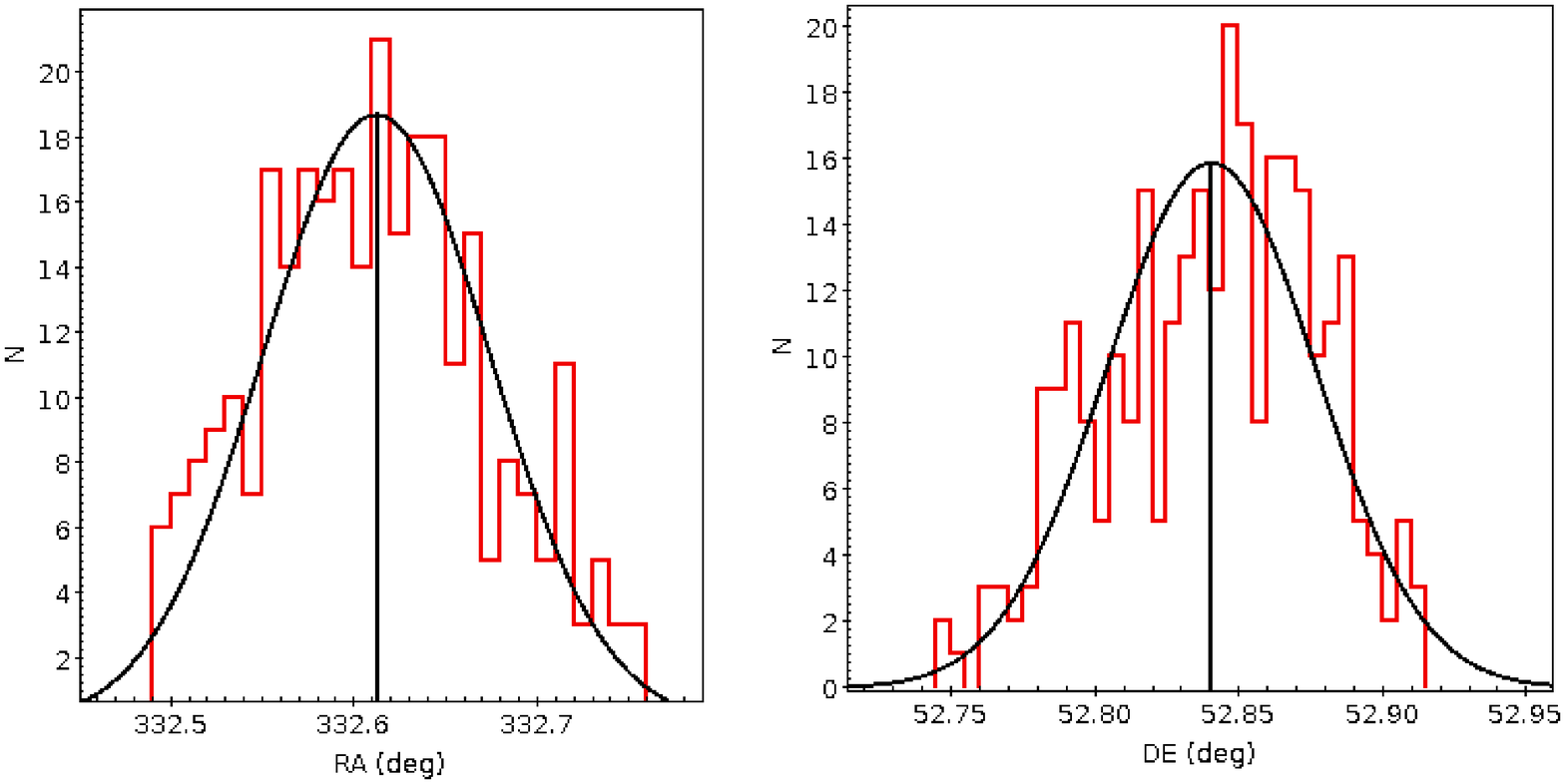}
\caption{Profiles of stellar counts across cluster IC 1434. The Gaussian fits have been applied. The center of symmetry about
the peaks of Right Ascension and Declination is taken to be the position of the cluster centers.
}
\label{center} 
\end{center}
\end{figure*}
%%%%%%%%%%%%%%%%

\subsection{Spatial structure:radial density profile}

The accuracy of central coordinates is very important for the reliable estimation of the cluster's main fundamental
parameters (e.g., age, distance, reddening, etc.). We applied a star-count technique to obtain the center coordinates
towards the area of IC 1434. The resulting histograms in both the RA and DEC directions are shown in the left panel
of Fig.~\ref{center}. The Gaussian curve-fitting at the central zones provides the center coordinates as
$\alpha = 332.612\pm0.06$ deg ($22^{h} 10^{m} 26.8^{s}$) and $\delta = 52.84\pm0.04$ deg ($52^{\circ} 50^{\prime} 24^{\prime\prime}$).
These estimated values are in very good agreement with the values given by Dias et al. (2002) and Cantat-Gaudin et al. (2018). 

\begin{figure*}
\begin{center}
\includegraphics[width=7.5cm, height=7.5cm]{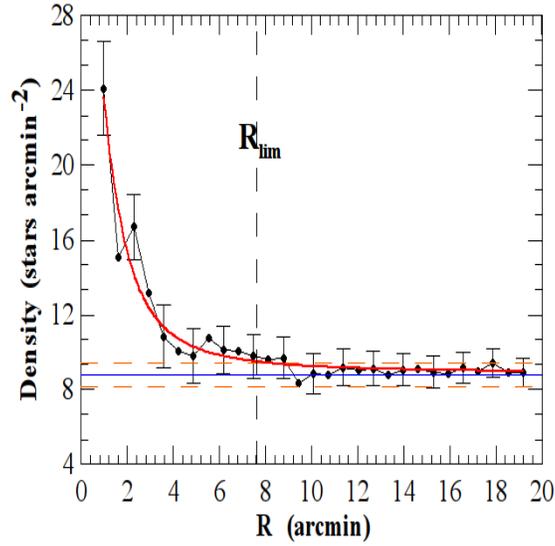}
\caption{Surface density distribution of the cluster IC 1434. Errors are determined from sampling statistics $\frac{1}{\sqrt{N}}$, where $N$ is the number of stars used in the density estimation at that point. The smooth line represents the fitted profile whereas
the dotted line shows the background density level. Long and short dash lines represent the errors in background density.}
\label{rdp} 
\end{center}
\end{figure*}

Radial density profile (RDP) has been plotted in Fig.~\ref{rdp} to estimate the radius of the cluster. We divided the
observed area of IC 1434 into several concentric rings. The number density, $R_{i}$, in the i$^{th}$ zone is determined by
using the formula $R_{i}$ = $\frac{N_{i}}{A_{i}}$, where $N_{i}$ is the number of stars and $A_{i}$ is the area of the i$^{th}$
zone. This RDP flattens at $R\sim$ 7.6 arcmin and begins to merge with the background density as clearly shown in the right panel
of Fig.~\ref{rdp}. Therefore, we consider 7.6 arcmin as the cluster radius. A smooth continuous line represents  the fitted King (1962) profile:\\

\begin{table}
\centering
\caption{Structural parameters of IC 1434. Background and central density is in the unit of stars per arcmin$^{2}$.
Core radius ($r_c$) is in arcmin.
}
\vspace{0.5cm}
\begin{center}
\small
\begin{tabular}{cccccc} 
\hline\hline
Name & $f_{0}$ &$f_{b}$& $R_{c}$&${c}$ & $\delta_{c} $ \\
\\
IC 1434 & $23.70\pm3.49$&$8.78\pm0.21$&$1.25\pm0.19$&$0.78$&$3.7$ \\
\hline
\end{tabular}
\label{stru_para}
\end{center}
\end{table}

%%%%%%%%%%%%%%%%
%%%%%%%%%%%%%%%%

%\begin{equation}
~~~~~~~~~~~~~~~~~~~~~~~~~~~~~~~~~$f(r) = f_{bg}+\frac{f_{0}}{1+(r/r_{c})^2}$\\
%\end{equation} 

where $r_{c}$, $f_{0}$, and $f_{bg}$ are the core radius, central density, and the background density level, respectively.
By fitting the King model to RDP, we estimated the structural parameters of IC 1434 are listed in Table 2.

We have estimated the concentration parameter using equation $c = log (\frac{r_{lim}} {r_{c}}$),
as given by Peterson \& King 1975). In the present study, the
concentration parameter is found to be 0.78 for IC 1434. Maciejewski \& Niedzielski (2007) reported that $R_{lim}$ may vary for
individual clusters from 2$R_{c}$ to 7$R_{c}$. The estimated value of $R_{lim}$ ($\sim$ 6.1$R_{c}$) shows a fair
agreement with the Maciejewski \& Niedzielski (2007). 
%We have estimated tidal radius of IC 1434 using the method described
%by Bisht et al. (2020b).
We obtained the density contrast parameter ($\delta_{c} = 1 + (f_{0}/f_{bg})$) for IC 1434 as 3.7. It is lower than
$7 \le \delta_{c} \le 23$ as derived by Bonatto \& Bica (2009). This demonstrates that IC 1434 is a sparse cluster.

\subsection{Age and distance estimation using CMD}

\begin{figure*}
\begin{center}
\includegraphics[width=12.5cm, height=12.5cm]{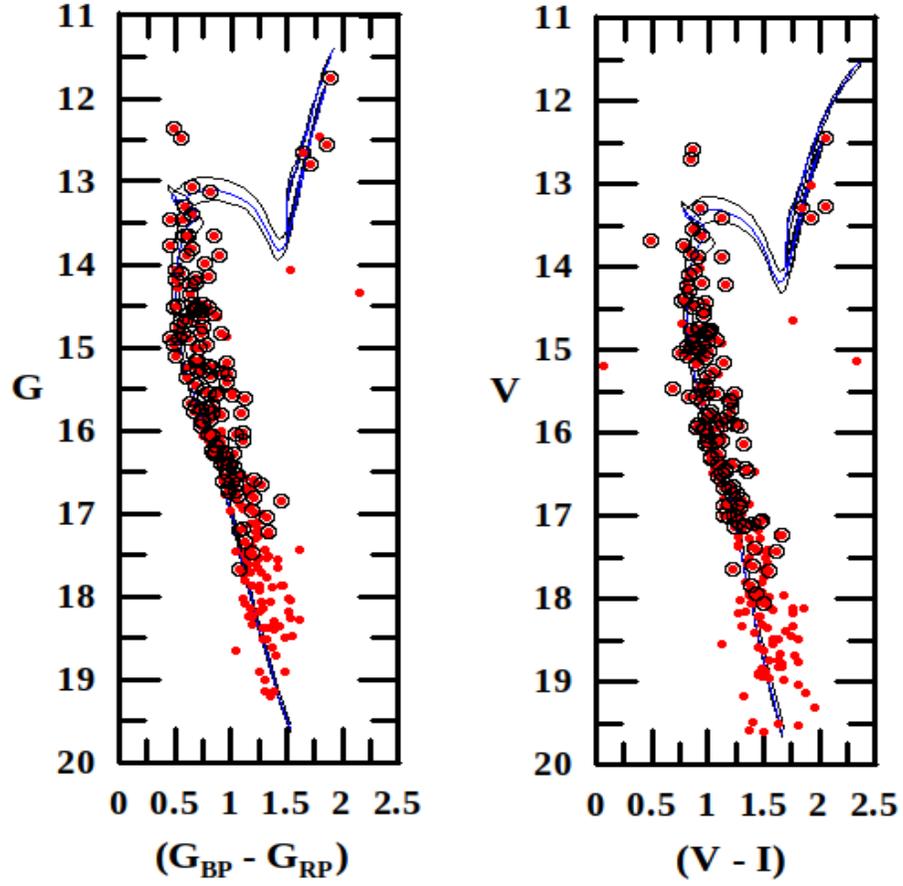}
\vspace{-0.5cm}\caption{Isochrone fitting to the CMDs. The curve is the solar metallicity isochrones taken from Marigo et al. (2017) of (log(age) = 8.75, 8.80, and 8.85). The filled and open circles are our 238 members and 127 members of Cantat-Gaudin et al. (2018), respectively.}
\label{cmd_fitting}
\end{center}
\end{figure*}

Age and distance are important parameters to trace the
structure and chemical evolution of the Milky Way Galaxy using OCs
(Friel \& Janes 1993).
The ($G$, $BP-RP$), and ($V$, $V-I$) color magnitude diagrams (CMDs) are shown in Fig.~\ref{cmd_fitting}. In this figure, filled dots
are  the most probable cluster members with membership probability $\ge$ 60$\%$ while open circles are matched ones with the catalog given by
Cantat-Gaudin et al. (2018). 
The age of IC 1434 has been estimated by fitting the theoretical isochrones of Marigo et al. (2017) with metallicity of $Z=0.0152$ to the CMDs as
shown in Fig.~\ref{cmd_fitting}. In this figure, we used the isochrones of different ages of log(age) = (8.75, 8.80 and 8.85). We found
the best global fit at log(age)=8.80, which is corresponding to $631\pm73$ Myr of cluster's age.

Our estimated value of color-excess in Gaia bands ($E(BP-RP)$) is 0.43 mag from the isochrones  fitting to the CMD's.
We have calculated
the interstellar reddening ($E(B-V)$) as 0.34 mag using the transformation equations ($E(B-V)$= 0.785 $E(BP-RP)$) as taken from
Abdelaziz et al. (2020). Distance modulus ($m-M_{G}$=12.51 mag) of IC 1434 provides the heliocentric distance as $3.2\pm0.1$ kpc.
Our estimated values of $E(B-V)$ and distance modulus are in fair agreement with the values $E(B-V)=0.49$ and $m-M_{G}$=12.43
as obtained by Angelo et al. (2020).

We obtained the Galactocentric distance as $9.4\pm0.1$ kpc by assuming 8.3 kpc distance of the Sun to the Galactic center.
The Galactocentric coordinates are estimated as $X_{\odot} = -0.55\pm0.04$ kpc, $Y_{\odot} = 3.1\pm0.2$ kpc and $Z{\odot} = -0.15\pm0.010$ kpc.
The estimated values of Galactocentric coordinates are very close to the values obtained by Cantat-Gaudin et al. (2018).

We also have checked the distance of IC 1434 using a parallax of stars from the Gaia~DR2 catalog. We found the distance of this object as 3.3 kpc which is very close to our estimation from the isochrone fitting method. Angelo et al. (2020), Cantat-Gaudin (2020),
Kharchenko et al. (2013), and Tadross (2009) have been determined the distance value of IC 1434 as 3.1, 3.3, 3.2, and 3.0 kpc,
respectively. Our derived value of distance is showing very good agreement with Angelo et al. (2020), Cantat-Gaudin (2020), and Kharchenko et al. (2013). The distance estimation of IC 1434 by Tadross (2009) is based on 2MASS and NOMAD data. Our obtained
value of the distance is much precise than Tadross (2009) because our estimation is based on good quality optical data along
with the high precision Gaia~DR2 astrometry.

%%%%%%%%%%%%%%%%
\section{Conclusions}
\label{con}

We presented the photometric and kinematic study of poorly studied open cluster IC 1434 using CCD $VRI$ and Gaia~DR2 data.
We have estimated the membership probabilities of stars towards the region of IC 1434 and have found 238 members
with a membership probability $\ge$ 60$\%$. We have used those probable members to derive the fundamental parameters.
The main results of the current investigation are
as follows:

\begin{itemize}

\item The updated cluster center coordinates are estimated as: $\alpha = 332.612\pm0.06$ deg ($22^{h} 10^{m} 26.8^{s}$) and $\delta = 
      52.84\pm0.04$ deg ($52^{\circ} 50^{\prime} 24^{\prime\prime}$) using the most probable cluster members. Cluster radii is obtained
      as 7.6 arcmin using radial density profile.\

\item On the basis of the vector point diagram and membership probability estimation of stars, we identified 238 most probable cluster members
      for IC 1434. The mean PMs of the cluster are estimated $-3.89\pm0.19$ and $-3.34\pm0.19$ mas yr$^{-1}$ in both the RA and DEC
      directions, respectively.\

\item Distance to the cluster IC 1434 is determined as $3.2\pm0.1$ kpc. This value is well supported by the distance estimated using mean
      parallax as 3.3 kpc. Age is estimated as $631\pm73$ Myr by comparing the cluster CMD with the theoretical isochrone given by
      Marigo et al. (2017) with $Z=0.0152$.\

\end{itemize}

\section{Acknowledgements}
The authors are thankful to the anonymous referee for useful comments, which improved the contents of the paper significantly.
This work is supported by the IMHOTEP collaboration program No. 42088ZK between Egypt and France. D. Bisht is supported by the Natural
Science Foundation of China (NSFC-11590782, NSFC-11421303). This work has made use of data from the European Space Agency (ESA) mission
Gaia (https://www.cosmos.esa.int/gaia), processed by the Gaia Data Processing and Analysis Consortium
(DPAC, https://www.cosmos.esa.int/web/gaia/dpac/consortium). Funding for the \\ DPAC has been provided by national institutions, in
particular the institutions participating in the Gaia Multilateral Agreement. This work has mad use the TOPCAT http://www.starlink.ac.uk/topcat. 
It has been developed mostly in the UK within various UK and Euro-VO projects (Starlink, AstroGrid, VOTech, AIDA, GAVO, GENIUS, DPAC) and
under PPARC and STFC grants. Its underlying table processing facilities are provided by the related packages STIL and STILTS.This research
has made use of Vizier catalogue access tool, CDS, Aladin sky atlas developed at CDS, Strasbourg Observatory, France. This work has made use
of data from the American Association of Variable Star Observers (AAVSO) Photometric All-Sky Survey (APASS) DR9 catalog.
%=======================================================================================

\section{References}
%\begin{thebibliography}
\begin{flushleft}
Abdelaziz A.E., Hendy Y.H.M., Shokry A., et al., 2020, Revista   Mexicana \\ \qquad de Astronomia y Astrofisica, 56, 245\\
Angelo M. S., Santos J. F. C., and Corradi W. J. B., 2020, MNRAS, 493, \\ \qquad 3473\\
Bisht D., Ganesh Shashikiran, Yadav R.K.S., et al., Geeta, 2018, \\ \qquad Adv. Space Res., 61, 517\\
Bisht D., Yadav R.K.S., Ganesh Shashikiran., et al., 2019, MNRAS, 482, \\ \qquad 1471B\\
Bisht D., Zhu Qingfeng., Yadav R.K.S., et al., 2020a, MNRAS, 494, 607-623\\
Bisht D., Elsanhoury W.H., Zhu Qingfeng, et al. 2020b, AJ, 160, 119B\\
Bonatto C., and Bica E., 2009, MNRAS 397, 1915\\
Cantat-Gaudin T., Jordi C., Vallenari A., et al., 2018, A\&A, 618, 93\\
Cantat-Gaudin T., Anders F., and Castro-Ginard A., 2020, A\&A, 640A, 1C\\
Chen L., Hou J.L., and Wang J.J., 2003, AJ, 125, 1397\\
%Chonis T.S., and Gaskell C.M., 2008, AJ, 135, 264\\
Dias W. S., Alessi B.S., Moitinho A, et al., 2002, A\&A, 389, 873\\
%Elmegreen B.G., 1999, ApJ, 515, 323\\
%Elmegreen B.G., 2000, ApJ, 539, 342\\
Friel, E. D., \& Janes, K. A. 1993, A\&A, 267, 75\\ 
Friel E.D., 1995, ARA\&A, 33, 381F\\ 
Gaia Collaboration et al., 2016a, A\&A, 595, A1\\
Gaia Collaboration et al., 2016b, A\&A, 595, A2\\
Gaia Collaboration et al., 2018a, A\&A, 616, A10\\
Gaia Collaboration et al., 2018b, A\&A, 616, A17\\
Gao Xin-hua, 2018, PASP, 130, 124101\\
Henden A., Munari U., 2014, Contrib. Astron. Obs. Skalnate Pleso, \\ \qquad 43, 518\\
%Heden A., Templeton M., Terrell D., et al. 2016, VizieR Online \\ \qquad Data Catalog, II/336\\
Heden A., Templeton M., Terrell D., et al. 2016, VizieR Online \\ \qquad Data Catalog, II/336\\
Hendy Y.H.M., 2018, NRIAG Journal of Astronomy and Geophysics, 7, \\ \qquad 180H\\
Janes K., and Adler D., 1982, ApJS, 49, 425J\\ 
Janes K.A., and Phelps R.L., 1994, AJ, 108, 1773J\\  
Jordi C., Gebran M., Carrasco J.M., et al., 2010, A\&A, 523, A48\\
Kharchenko N.V., Piskunov A.E., Roser S., et al., 2004, Astron. Nachr., 325, \\ \qquad 740\\
Kharchenko N.V., Piskunov A.E., Schilbach E., et al., 2013, A\&A, 558, A53\\
King I., 1962, AJ, 67, 471\\
%Kroupa Pavel, 2011, IAUS, 270, 141K\\
%Kroupa, Pavel, 2001, MNRAS, 322, 231K\\ 
Marigo P., Girardi L., Bressan A., et al., 2017, ApJ, 835, 77\\
Maciejewski G., and Niedzielski A., 2007, A\&A, 467, 1065\\
Maciejewski G., and Niedzielski A., 2008, Astron. Nachr., 329, 602\\ 
Meynet G., Mermilliod J.-C., and Maeder A., 1993, A\&AS, 98, 477M\\
Monet David G., Levine Stephen E., Canzian Blaise, et al., 2003, AJ, 125, \\ \qquad 984M\\ 
Peterson C.J., and King I.R., 1975, AJ, 80, 427\\
%Piskunov A. E., Belikov A.N., Kharchenko N. V., et al., 2004, MNRAS, \\ \qquad 349, 1449\\
Phelps R. L., and Janes K.A., 1993, AJ, 106, 1870\\
Rangwal G., Yadav R. K.S., Durgapal A., et al., 2019, MNRAS, 490, 1383\\
%Richtler Tom, 1994, A\&A, 287, 517R\\
Salgado J., Gonzalez-Nunez J., Gutierrez-Sanchez R., et al., 2017, A\&C, 21, \\ \qquad 22S\\
Salpeter E.E., 1955, ApJ, 121, 161\\
Sandage A., 1988, BAAS, 20, 1037S \\
%Scalo J.M., 1986, Fund. Cosmic Phys, 11, 1\\
Spitzer L., and Hart M., 1971, ApJ, 164, 399\\
Stetson P.B., 1980, AJ, 85, 387\\
Stetson P.B., 1987, PASP, 99, 191\\
Stetson P.B., 1992. In: Warrall, D.M., Biemesderfer, C., Barnes J., (Eds.), \\ \qquad Astronomical
Data Analysis Software and System I. ASP Conf. Ser. \\ \qquad vol. 25, Astron. Soc. Pac., San
Francisco, pp. 297\\
Tadross A. L., 2009, New Astronomy, 14, 200\\
Tadross A.L., and Hendy Y.H.M., 2016, Journal of the Korean  \\ \qquad Astronomical Society, 49, 57\\
Tadross A.L., Bendary R., Hendy Y., et al., 2018, Astron. Nachr., 339, \\ \qquad 698\\
Twarog Bruce A., Ashman Keith M., and Anthony-Twarog Barbara J.,\\ \qquad 1997, AJ, 114, 2556T\\ 
Vasilevskis S., Klemola A., and Preston G., 1958, AJ, 63, 387\\
%Yadav R.K.S., and Sagar R., 2002, MNRAS, 337, 133\\
%Yadav R.K.S., and Sagar R., 2004a, MNRAS, 349, 1481\\
Yadav R.K.S., Sariya D.P., and Sagar R., 2013, MNRAS, 430, 3350\\
Zhao J.L., and He Y.P., 1990, A\&A, 237, 54\\
%Zhao J. L., and Shao Z. Y., 1994, A\&A, 288, 89\\
%\end{thebibliography}
\end{flushleft}
\end{document}